# Highly effective gating of graphene on GaN

Jakub Kierdaszuk [1*], Ewelina Rozbiegała [2,3], Karolina Piętak [2,4], Sebastian Złotnik [2,5], Aleksandra Przewłoka [2,6,7], Aleksandra Krajewska [2,6], Wawrzyniec Kaszub [2], Maria Kamińska [1], Andrzej Wysmołek [1], Johannes Binder [1], Aneta Drabińska [1]

[1] *Faculty of Physics, University of Warsaw, Poland*
[2] *Łukasiewicz Research Network - Institute of Microelectronics and Photonics, Centre for Electronic Materials Technology, Warsaw, Poland*
[3] *Warsaw University of Technology, Faculty of Materials Science and Engineering,*
[4] *Warsaw University of Technology, Faculty of Chemistry, Poland*
[5] *Military University of Technology, Institute of Applied Physics, Warsaw, Poland*
[6] *Center for Terahertz Research and Applications (CENTERA), Warsaw, Poland*
[7] *Institute of Optoelectronics, Military University of Technology, Warsaw, Poland*

*Corresponding author. Tel. +48 573931438, E-mail: jakub.kierdaszuk@fuw.edu.pl

Abstract

By using four-layered graphene/gallium nitride (GaN) Schottky diodes with an undoped GaN spacer, we demonstrate highly effective gating of graphene at low bias rendering this type of structure very promising for potential applications. An observed Raman G band position shift larger than 8.5 cm$^{-1}$ corresponds to an increase in carrier concentration of about $1.2 \cdot 10^{13}$ cm$^{-2}$. The presence of a distinct G band splitting together with a narrow symmetric 2D band indicates turbostratic layer stacking and suggests the presence of a high potential gradient near the Schottky junction even at zero bias. An analysis based on electroreflectance measurements and a modified Richardson equation confirmed that graphene on n-GaN separated by an undoped GaN spacer behaves like a capacitor at reverse bias. At least 60% of G subband position shifts occur at forward bias, which is related to a rapid reduction of electric field near the Schottky junction. Our studies demonstrate the usefulness of few layer turbostratic graphene deposited on GaN for tracing electron-phonon coupling in graphene. Multilayer graphene also provides uniform and stable electric contacts. Moreover, the observed bias sensitive G band splitting can be used as an indicator of charge transfer in sensor applications in the low bias regime.

1. Introduction

In recent years, much effort has been devoted to the development of graphene and its implementation in nanodevices. Graphene-based top electrodes that could revolutionise light emitting diodes[1,2] due to its high transparency[3], conductivity[4] and mechanical durability[5] raised great hope for application. Furthermore, variable barrier graphene/semiconductor devices with high gating efficiency, which are fundamental building blocks for the construction of more complex systems like solar cells or optical sensors, were reported.[6–10] Heterostructures of wide band gap gallium nitride (GaN) covered with graphene are promising for fabrication of effective optoelectronic devices.[11–14] Because direct growth of graphene on GaN results in highly defective material, it becomes necessary to transfer graphene from different substrates. The transfer of graphene from copper onto GaN results in high-quality, large area graphene layers.[15–17] Nevertheless, graphene on copper is polycrystalline and differently oriented grain boundaries can negatively impact its properties.[18] Moreover, it has been shown that small islands of bi- or three-layer graphene can also be present as well as point defects.[19] Due to the mechanical vulnerability of monolayer graphene, few layer graphene is more suitable for these structures. It allows to obtain large area sheets with a low number of cracks and due to the increased stability it enables to establish a more stable electrical contact with a lower resistance compared to monolayer graphene.[1,20] The Schottky barrier height in a graphene/GaN junction is reported to be in the range from 0.33 eV to 0.9 eV and therefore effects like rectification or gating are present.[6,21,22] An implementation of appropriate experimental techniques to investigate the properties of these kind of heterostructures is crucial for their future application. One way to trace a gating effect is optically by Raman spectroscopy which is a non-destructive and versatile tool to study graphene.[23,24] Due to the interaction between lattice vibrations and carriers, the shapes and energies of the two main graphene bands, G (around 1583 cm$^{-1}$) and 2D (around 2678 cm$^{-1}$), depend on both, strain and carrier concentration.[23–25] The charge carrier density of graphene in gated structures is modified by an applied electric field and can be traced by a detailed peak analysis of the Raman bands (G and 2D band energies, full width at half maximum (FWHM) of G band and ratio of 2D and G band intensities $R_{2DG}$).[26,27] The most common kind of gated structures is based on a capacitor geometry, i.e. graphene/dielectric/conductive semiconductor (e.g. graphene on top of SiO$_2$/deposited on doped silicon).[26,27] In this geometry graphene forms the top electrode and a bottom electric contact is made to the doped semiconductor (or metal). This type of structures were previously used for studies of one, two or three layers of graphene grown on copper and transferred onto SiO$_2$/Si substrates.[19] However, a large thickness of the dielectric layer (typically 90 nm - 300 nm) requires a large applied voltage of up to 100 V.[19,26] A solution-gate field effect transistor, in contrast, uses the ultra-thin electronic double layer of an ideally polarizable electrode at the graphene/solution interface allowing for much larger gating efficiency.[28–31] This system features extremely high capacitances and lower applied voltages and is therefore promising for future nanosensing applications. However, the interaction between graphene and the electrolyte can also enhance degradation of graphene by electrochemical formation of radicals which can attack the graphene electrode.[32]

Here, we demonstrate highly effective charge carrier modulation obtained in a GaN-based Schottky diode structure with an undoped GaN spacer. Previous studies by S. Tongay et al. show promising perspectives of graphene/semiconductors Schottky diodes in the reverse biased regime for several doped semiconductors like Si, GaAs, 4H-SiC and GaN.[6] In our work graphene is measured in situ as part of a Schottky diode rather than a gated field effect transistor.

Here, we also study the forward bias direction for which we observe much larger changes in Raman spectra.

In this work 4-layer graphene (4Lgr) with random layer stacking was used as a top electrode on a GaN epitaxial platform. I-V characteristics were measured to electrically characterize the obtained diode while Raman spectroscopy measurements were applied to study the evolution of the Raman spectra as a function of bias. Furthermore, we performed electroreflectance measurements, which show good agreement with the theory of graphene/semiconductor Schottky diodes. We demonstrate that GaN-based diodes with a graphene electrode and an undoped GaN spacer can induce a large gradient of carrier concentration between the consecutive layers in multilayer graphene. This observation provides new opportunities concerning studies of electron-phonon interaction in graphene and other phenomena related to interlayer interactions in few-layer graphene on GaN platform.

2. Experimental details

2.1. Sample preparation

The GaN epilayer structure was grown by Metalorganic Chemical Vapour Deposition on a c-plane sapphire substrate using a AIX 200/4 RF-S MOVPE system. A 750 nm thick AlN buffer layer proceeded by a 300 nm thick layer of undoped GaN and a 1.2 µm thick layer of GaN doped with silicon (n-type, electron concentration up to $4 \cdot 10^{18}$ cm$^{-3}$) was grown on (0001)-oriented sapphire (Al$_2$O$_3$) wafers. Finally, the layers sequence was finished with a 100 nm thick layer of undoped GaN (electron concentration around $2 \cdot 10^{16}$ cm$^{-3}$).

Graphene was grown on copper foil by a standard CVD (chemical vapor deposition) method.[16] With this growth method, local inclusions of bilayer graphene (results of primary CVD growth) on the copper surface can be present. 4Lgr was prepared in three steps. Firstly, a graphene monolayer grown on a copper substrate was transferred onto another graphene monolayer on a copper substrate by high-speed electrochemical delamination method.[16]

Then, a third graphene monolayer was transferred on the two layers of graphene obtained in the previous step. Finally, a fourth graphene monolayer was transferred on the graphene trilayer obtained in the previous step. With this process one obtains a mechanically stable 4Lgr, however, the so fabricated multi-layer graphene has an inhomogeneous and uncontrolled layer stacking. Then, a polymer frame method was used to transfer the 4Lgr from copper onto the aforementioned GaN platform.[33] Graphene layers were placed in the centre of the sample surface in order to avoid short circuits. The whole area of the graphene electrode was 0.25 cm$^2$. Two ohmic indium contacts to the conductive GaN layer were made and verified by I-V measurement with a resistance around 260 $\Omega$, while silver paint was used for the graphene contact. A schematic drawing of this structure together with the I-V measurement scheme (an Agilent B2901A SMU was used for the measurements) are presented in Fig. 1. The same electrical setup was also used for applying a constant voltage during optical measurements.

2.2. Characterization technique

Raman spectra were measured with a Renishaw InVia spectrometer equipped with x100 objective and 532 nm laser excitation. Excitation power was reduced to several mW to minimize heating effects. Raman micromapping was performed on an area of several square micrometers with 100 nm step. Electroreflectance (ER) measurements are realized by small modulation of a constant voltage applied to the sample. The benefit of this technique is that it gives very sharp, sensitive to build-in electric field, lines even at room temperature. The reflectivity was measured with the use of a monochromatic light obtained from 75 W Xe lamp. The DC and AC signal was measured by a voltmeter and a lock-in amplifier, respectively.

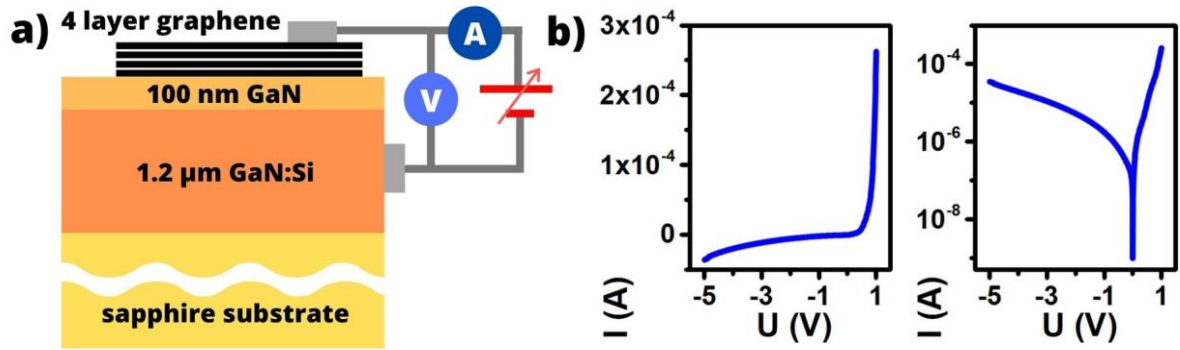

Figure 1. a) Schematic drawing of the GaN / graphene heterostructure indicating the four graphene layers and the electrical measurement setup, b) sample I-V characteristic in linear and logarithmic scale.

3. Results and discussion

I-V measurements clearly show diode characteristics for the graphene-GaN junction (Fig. 1 b). At 1 V of forward bias the current value exceeds $10^{-4}$ A while in reverse bias direction current varies from $10^{-9}$ A at 0 V to $10^{-5}$ A at –5 V (Fig. 1b). The current density in our sample at –5 V (around $10^{-4}$ Acm$^{-2}$) is similar to results reported by Tongay et al. and four orders of magnitude higher than observed in small area lithography made graphene/GaN Schottky contacts.[6,22] A fit to the I-V curve of the sample allows to extract a Schottky barrier height of 0.8 eV, which is in good agreement with other experimental results.[21,22] The presence of a significant reverse current might be related to the bias induced shift of Fermi energy in graphene which will be discussed in further paragraphs.[6,34]

Raman spectra were measured as a function of bias from 1 V to −5 V and back with a 0.5 V step. Each Raman spectrum was measured with the same laser power and acquisition time. A detailed analysis of the G band evolution (Fig. 2) showed three different types of spectra (called type A, B and C) with one, two or three G subbands, respectively (Fig.2a). An additional figure showing the fitted spectra is presented in the Supplementary Materials. According to their energy, in this paper, G subbands will be referred to as G1 (about 1580 cm$^{-1}$), G2 (about 1589 cm$^{-1}$) and G3 (about 1598 cm$^{-1}$). A large splitting of the G band suggests the presence of large gradient of electric field near graphene/GaN junction, which will be discussed in further paragraphs. Interestingly, the 2D band generally consists of a single subband for all measured spectra. Evolution of the B-type spectrum for selected biases is shown in Fig. 2b and c. Generally, a strong redshift of the G band position is observed with increasing bias, (Fig. 2b) while the 2D band position is nearly constant (Fig. 2c).

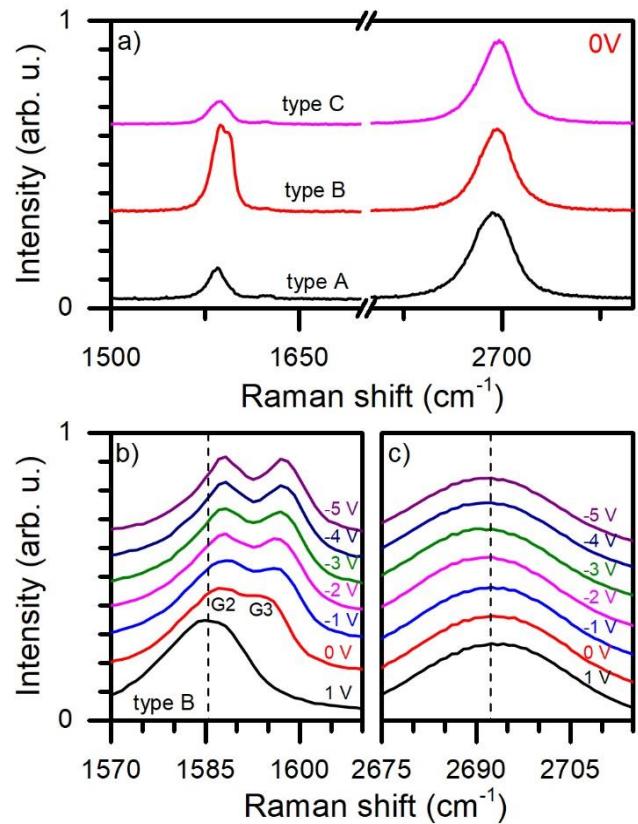

Figure 2. Raman spectra measured under 532 nm excitation at different biases. a) three representative types of spectra measured at 0 V, b) G band evolution for the B-type spectrum, c) 2D band evolution for the B-type spectrum.

Table 1. G band position shifts (ΔE), G energies, G and 2D FWHM at 0 V and (in cm$^{-1}$) for three types of spectra in four layer graphene at GaN heterostructure.

| Spectrum type | Subband | ΔE (1 V → -1 V) | ΔE (-1 V → -5 V) | G Energy | G FWHM | 2D FWHM |
|---|---|---|---|---|---|---|
| A | G1 | 1.8 | 0.3 | 1584.5 | 16.7 | 37.3 |
| B | G2 | 5.1 | 0.3 | 1586.5 | 13.8 | 29.9 |
|   | G3 | 8.5 | 1.1 | 1594.9 | 6.9 |  |
| C | G1 | 1.6 | 0.8 | 1580.4 | 12.6 | 31.7 |
|   | G2 | 3.2 | 1.4 | 1587.3 | 10.4 |  |
|   | G3 | 6.6 | 1.4 | 1593.7 | 8.4 |  |

Each G subband was fitted by a Lorentzian function. The evolution of G band positions as a function of bias is presented in Figure 3. The most significant redshift in position of both G band components occurs in the range between 1 V and –1 V (Fig. 3a, Table 1). It varies from 1.8 cm$^{-1}$ in A-type spectrum to 8.5 cm$^{-1}$ for the G3 component in the B-type spectrum. A further small increase of the G subband positions in reverse bias direction (up to 1.4 cm$^{-1}$) has a linear behavior. Thus, 70-90% of G subband position shifts occur in the range between -1 V and 1 V. Moreover, the largest shifts are characteristic for the subband with the highest energy and the lowest FWHM (Table 1). The energetic difference between the subbands also increases as a function of bias. For type-B spectra this value increases from 5.9 cm$^{-1}$ at 1 V to 10.1 cm$^{-1}$ at –5 V. This corresponds to the carrier concentration gradient increase from $8·10^{12}$ cm$^{-2}$ to $1.4·10^{13}$ cm$^{-2}$ (recalculated by formula proposed by [35]). Similar trends are present for type–C spectra. The analysis of the whole sweep cycle shows the presence of small hysteresis which, however, does not change the overall trend (Supplementary Materials). This hysteresis might be related to charging of deep defect states in GaN which results in the creation of local internal electric field in GaN or due to surface states at the graphene / GaN interface.

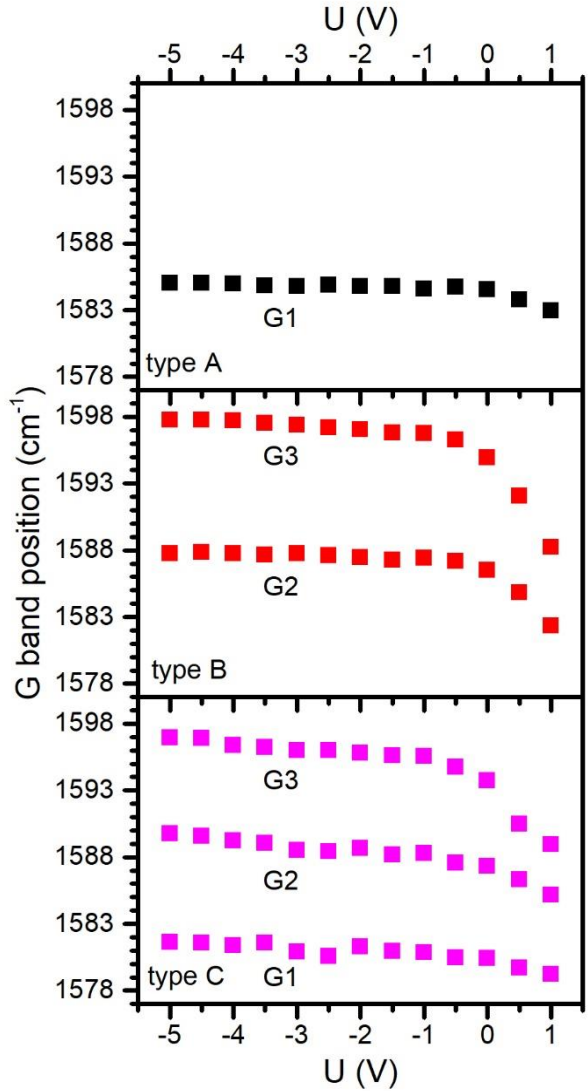

Figure 3. Dependence of the G band position on bias for the three types of spectra A, B, C. G1, G2, and G3 indicate different subbands of the graphene Raman G band. The measurement B correspond to the spectra shown in Figure 2b.

Generally, for undoped monolayer graphene the G band energy has the lowest value.[26] Due to the presence of a Kohn anomaly near the Γ point of the graphene Brillouin zone, an increasing G band position is observed for hole and electron doping. Therefore, a V-shape model dependence of the graphene G band energy on carrier concentration is observed.[23] The phonon lifetime increases for larger carrier concentrations and due to the uncertainty principle a decrease of G band FWHM is observed.[23,26,27] Therefore, the G band FWHM as a function of doping level exhibits the opposite behaviour than the G band energy and the maximum value is observed for undoped monolayer and few layer graphene.[19]

In our studies few layer graphene with different stacking was used as top electrode. In this type of graphene, layers can behave like independent monolayers with larger distances compared to AB stacking and weak interlayer interactions.[36,37] Moreover, graphene layers were assembled manually without a high temperature annealing step. Therefore, interlayer distances might by higher than in exfoliated samples of few layer graphene. This assumption has been confirmed by the analysis of 2D band FWHM. In our measurements it varies from 30 cm$^{-1}$ to 37 cm$^{-1}$ which is characteristic rather for monolayer graphene than 4 layer graphene.[38,39] This value is also significantly lower than reported for trilayer graphene with ABA or ABC stacking (40-60 cm$^{-1}$ or more than 60 cm$^{-1}$ respectively).[40–42] The absence of significant 2D subbands indicate that interlayer interactions in our case should be weak or absent and should not impact the graphene Raman band evolution as it is the case for few layer graphene with high stacking symmetry.[19] Therefore, the G band splitting cannot be related to optical phonon mixing and the breaking of inversion symmetry, as it was reported in bilayer graphene with AB stacking.[43,44]

In relation to the trends presented in Figure 3, a high electron doping is present. Values of maximum G subband position shift obtained in our gating experiment are slightly higher than previously reported position shift for trilayer graphene (6.9 cm$^{-1}$ at 50 V)[19], for bilayer graphene (3.6 cm$^{-1}$ at 1 V)[44] and results for monolayer graphene (6.5 cm$^{-1}$ at 100 V).[26] Thus, our observation suggest a large modification of electric field near the diode junction even at low bias. Furthermore, a negative correlation between the G subband energies and their FWHMs are observed (Table 1). This excludes the presence of a strain gradient for which a positive correlation between G band energy and its FWHM is commonly observed.[24] This result suggests that similarly to other works, the subband with the largest energy is characterized by the highest value of carrier concentration which confirms the presence of a large potential gradient across the graphene layers.[44] Different number of G subbands might be related to the local number of graphene layers arising due to the damages caused by transfer method. Local differences in interlayer spacing or stacking as well as low or similar doping level of some layers might also affect the number of visible subbands. The assessment of the local stacking and interlayer distance requires measurement techniques on the nanometer and angstrom scale, which is, however, beyond the scope of this paper.

The analysis of two other parameters: the G band FWHM and the intensity ratio between 2D and G band ($R_{2DG}$) and their evolution as a function of bias also confirms the trends presented for the G band position (see Supplementary Materials). For backward bias, the G band FWHM decreases for each subband. A similar behavior as well as a hysteresis is observed for the $R_{2DG}$ parameter. Results of a two-dimensional micromapping show that the gating effect is strong and present all over the measured surface (see Supplementary Materials). Such a small variation of gating efficiency is promising for future control of this phenomena which might be useful for the construction of gated devices.

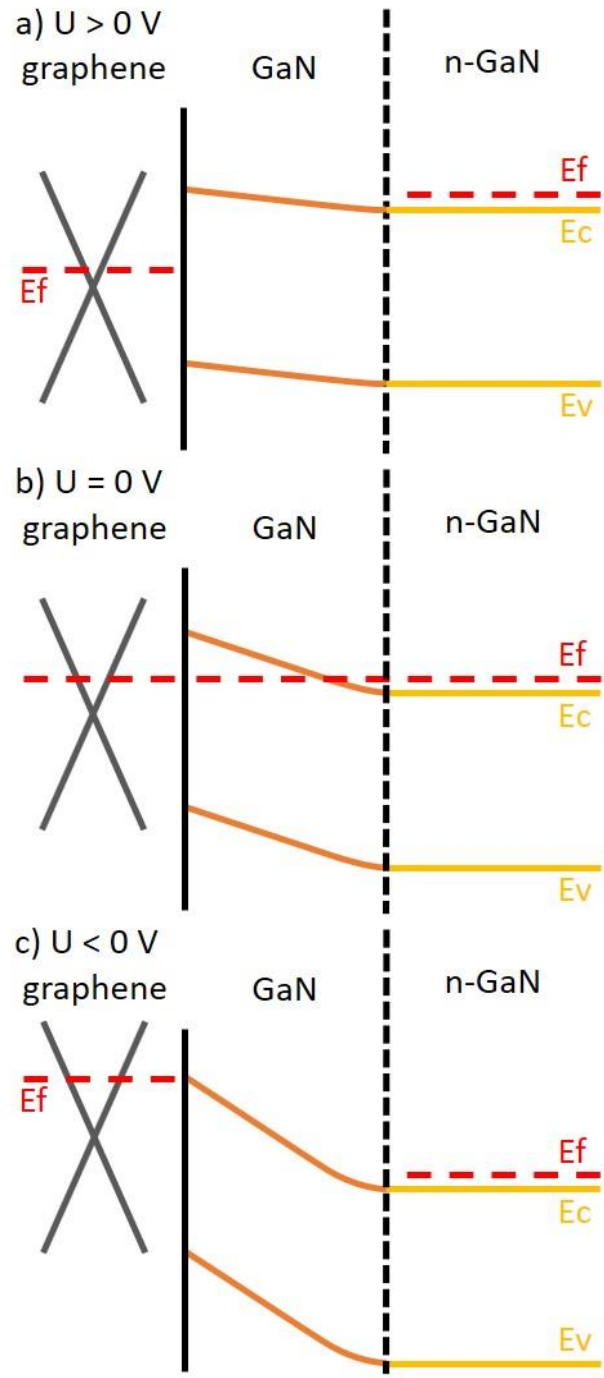

Figure 4. Band structure of graphene/GaN/n-GaN Schottky diode and Fermi level position for: a) forward bias, b) zero bias, c) reverse bias.

In order to illustrate the behaviour of our diode and the role of the GaN 100 nm spacer a band diagram of the diode structure is presented in Fig. 4. The band alignment in graphene/GaN Schottky diodes with a n-doped GaN layer suggests that for forward (backward) bias the Fermi level $E_F$ of graphene should decrease (increase).[6] Therefore, applying a reverse (forward) voltage should increase (decrease) the number of electrons in n-type doped samples. In the work of Tongay S. et. al. shifts under reverse bias were observed, however, the authors did not observe any significant changes for Raman bands at forward bias.[6]

For simplification, we present only one graphene layer in the diagram in Figure 4. The carrier concentration in undoped GaN spacer is two orders of magnitude lower than in the n-doped GaN. Therefore, at zero bias conduction and valence bands of GaN spacer have linear dispersion with a small bend at the interface with n-GaN (Fig. 4b). Even without any applied voltage, the interaction between GaN and graphene induces a strong charge transfer, so $E_F$ in graphene is high and n-type doping is present. The presence of high potential gradient is confirmed by the presence of G band splitting even at 0 V. At reverse bias (Fig. 4c), the graphene bands move up while conduction and valence bands of GaN move down. The linear slope of the band across the GaN spacer increases. In this regime, amount of free carriers in GaN is low and the whole structure behaves like a capacitor. Therefore, the $E_F$ of graphene depends only on the value of the electric field induced between n-GaN and the graphene electrode and changes less rapidly than around 0 V. At forward bias, graphene and n-GaN bands move down and up respectively (Fig. 4a). The slope of the band of the GaN spacer also decreases. Consequently, the situation established though charge transfer between n-GaN and graphene without an applied voltage is changed, which results in a rapid decrease of electron concentration in graphene.

A schematic spatial distribution of carrier concentration within subsequent graphene layers is shown in Figure 5. Initially, four-layer graphene has a large gradient of electron concentration. The highest value is present near the GaN surface. An applied reverse bias further increases the carrier concentration and its gradient and hence the splitting of the subbands. The most significant modification of carrier concentration occurs at about 0 V. In forward bias direction, the carrier concentration across subsequent layers becomes more uniform so the band splitting is less significant. The observed bias switchable carrier concentration gradient in few layer turbostratic graphene shows great potential for the development of sensors with the Raman spectrum being very sensitive to small changes in the forward voltage direction.

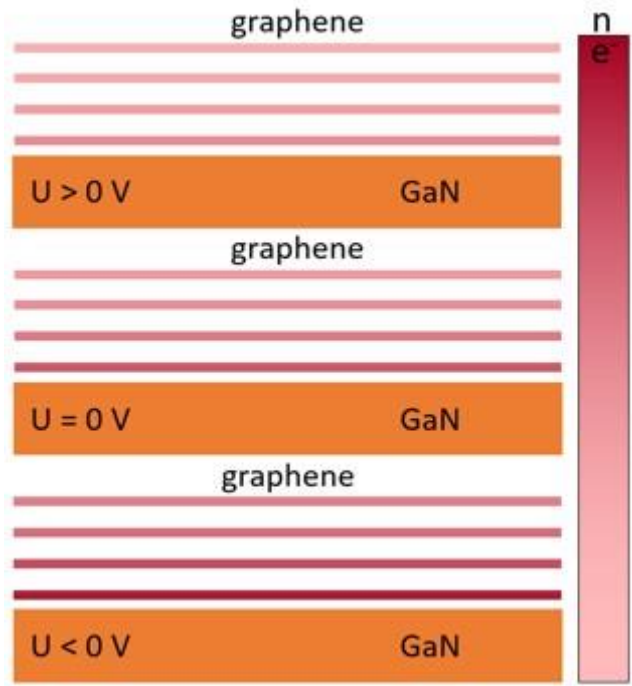

Fig 5. Scheme of graphene carrier concentration at different bias regimes. The colours of the graphene layers correspond to different values of electron carrier concentration.

The analysis of the band alignment for our sample and the G band evolution suggested that our structure behaves like a capacitor for backward bias. In order to estimate how the carrier concentration is changed by the applied electric field, electroreflectance measurement as a function of bias was performed (Fig. 6). Electroreflectance is a very powerful technique for the characterization of optical properties of semiconductors and their heterostructures.[20,45–48] These measurements enable to estimate energies of optical transitions, concentration of polarization induced charges and depletion width. A modulation of an electric field affects the dielectric function of GaN which change the reflection of the sample. For the case of a medium electric field, Franz-Keldysh oscillations appear in the reflectance spectrum, which in asymptotic approximation are described by the equation:[49],[50]

$$\frac{\Delta R}{R}(E) \sim \frac{1}{E^2(E-E_G)} e^{-\Gamma\sqrt{\frac{E-E_G}{(\hbar\Omega)^3}}} \cos\left(\frac{2}{3}\sqrt{\left(\frac{E-E_G}{\hbar\Omega}\right)^3} + \phi\right), \qquad (1)$$

where the position of the signal is determined by its energy band gap ($E_G$), while its period by an electrooptical energy ($\hbar\Omega$). The electrooptical energy depends on the electric field ($F$) and the concentration of induced charges ($n$) (eq. 2), where $\varepsilon$ and $\varepsilon_S$ are permittivity constants of vacuum and GaN, respectively.[49]

$$(\hbar\Omega)^3 = \frac{(e\hbar F)^2}{8\mu_\parallel} = \frac{1}{8\mu_\parallel}\left(\frac{e^2\hbar n}{\varepsilon\varepsilon_S}\right)^2, \qquad (2)$$

Contactless electroreflectance technique was previously used for studies of $E_F$ variation in GaN/graphene Schottky diode.[51] However, the fabrication of electric contacts enables us to measure ER as a function of bias (Fig. 6a, b) and further to estimate the dependence of the electric field and the concentration of surface carriers on bias (Fig. 6c). The extracted value of carrier concentration varies linearly from $5 \cdot 10^{11}$ cm$^{-2}$ at 0 V to $2.4 \cdot 10^{12}$ cm$^{-2}$ at -5 V which is characteristic for a capacitor-like behaviour. The change in concentration corresponds to an upshift of the graphene Fermi level by 112 meV. Electroreflectance results are compared with Raman measurements. The G band evolution was recalculated by a simple dependence of the G band energy on carrier concentration which is linear for small carrier concentrations.[35] However, the G band energy depends also on strain. This component was estimated by considering the 2D band position shift. It should not depend on carrier concentration in the medium electron doping regime and therefore the 2D band position shift is proportional to the G band strain induced position shift.[35] After taking into account the values of 2D and G band for unstrained and undoped graphene (2678 cm$^{-1}$ and 1580 cm$^{-1}$ respectively, and strain coefficient $\Delta E_G / \Delta E_{2D}$ equal to 0.4)[35], the G band position shifts for both subbands were recalculated into shifts of $E_F$ in graphene.[19,23,24,35]

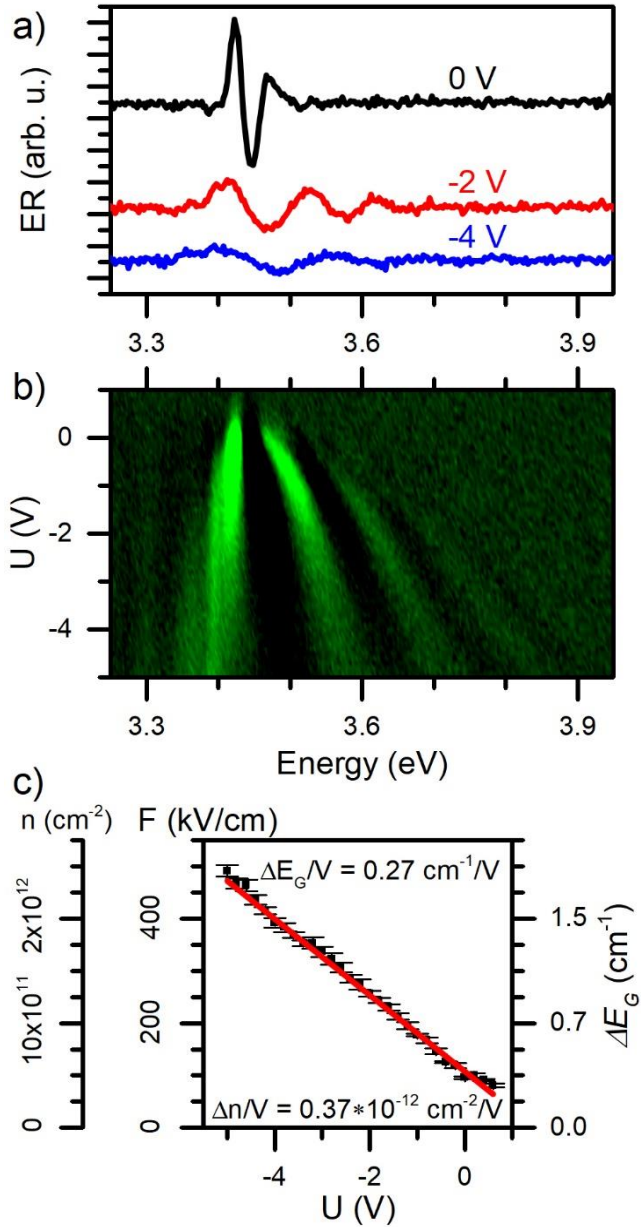

Figure 6. a) Electroreflectance spectra measured at different biases, b) electroreflectance map as a function of bias, c) bias dependence of the electric field induced in GaN. The additional axes show the conversion of the electric field to corresponding carrier concentrations (parallel plate capacitor model) and position shift of the G band in graphene.

$E_F$ shift with bias from IV characteristic was calculated using the Richardson model with a modification of the current density (as proposed by Tongay S., et al. taking into account the contribution of a Fermi level shift of graphene to the reverse current), described by equation 3, where $\phi_{SBH}$ is the Schottky barrier height and $A^*$ the Richardson constant. $A^*$ characterizes the thermionic emission from semiconductors and depends on the effective mass of the electron. Taking into account an effective electron mass of GaN equal to $0.22 m_e$, a value of Richardson constant equal to 26.4 Acm$^{-2}$K$^{-2}$ was obtained.[22,52,53].

$$J(T,V) = A^* T^2 e^{-\frac{e\varphi_{SBH} - \Delta E_F(V)}{k_B T}} \left( e^{\frac{eV}{k_B T}} - 1 \right), \tag{3}$$

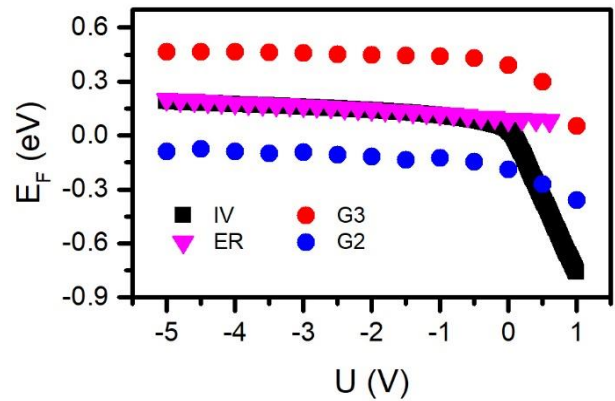

Figure 7. Dependence of the recalculated Fermi level ($E_F$) shift as a function of bias obtained from: I-V characteristic (recalculated from eq. 3, black squares), ER results (surface carrier concentration, magenta triangles), G band from B-type spectrum (red and blue circles).

However, this model is phenomenological and explains only some particular experimental features observed in previous experiments.[7] Nevertheless, Figure 7 shows good agreement for the Fermi level shift extracted from the I-V characteristics, Raman spectroscopy and electroreflectance measurements for the reverse bias. This confirms that the undoped GaN spacer between n-GaN and graphene at reverse bias has indeed a low carrier concentration and becomes highly insulating, so that the whole structure behaves like capacitor. The G subbands show a similar dependence except, that the Fermi energy calculated from G2 band is negative (Fermi level is shifted below the Dirac point) which disagrees with the result from the Raman evolution. This discrepancy is related to difficulties in estimation of the value of strain and G band energy for undoped few-layer graphene on GaN.

In our case it was difficult to apply larger forward bias due to the danger of diode destruction at high forward current. Studies of graphene at GaN with different doping might enable to reach the G band energy at neutrality point. The clear change of the Raman G band between 0 V and 1 V is a particularly good indicator and could be used to establish an optical readout of charge transfer via Raman spectroscopy of a turbostratic few layer graphene / GaN diode.

## 4. Conclusions

We have shown that graphene/GaN heterostructure based Schottky diodes show excellent performance in graphene gating even with low applied biases (on the order of hundreds of millivolts). A detailed analysis of the G band position shift enabled us to estimate that four-layer graphene deposited on n-doped GaN is also n-doped. Furthermore, a clearly visible G band splitting was observed and studied as a function of bias. The presence of splitting into two or three subbands suggests that the phenomena is related to the large gradient of electric potential near the graphene/GaN junction. G subbands behave like independent graphene layers. Moreover, the local potential gradient is large enough to observe this effect even without bias. In contrast to solution gated structures, GaN Schottky diodes do not require any additional medium to obtain similar effectiveness. The largest G band position shift occurs at forward bias. An application of few layer turbostratic graphene on GaN, as shown in this work, enables the construction of gated devices and sensors. Such systems may be used not only for precise measurements of G band position shift but also for tracing the clearly visible G band splitting. This effect has been identified as the fingerprint of effective gating in turbostratic graphene, which can be useful in characterization of other devices based on graphene/GaN heterostructures and opens new pathways for the construction of nanosensors.

5. Acknowledgements

This work was partially supported by the Ministry of Science and Higher Education in years 2015-2019 as a research grant "Diamond Grant" (No. DI2014 015744), the National Science Centre, Poland grant 2014/13/N/ST3/03772 and the Research Foundation Flanders (FWO) under Grant no. EOS 30467715.

Karolina Piętak acknowledges financial support from IDUB project (Scholarship Plus programme).

6. Bibliography

[1] D.-W. Jeon, W.M. Choi, H.-J. Shin, S.-M. Yoon, J.-Y. Choi, L.-W. Jang, I.-H. Lee, Nanopillar InGaN/GaN light emitting diodes integrated with homogeneous multilayer graphene electrodes, J. Mater. Chem. 21 (2011) 17688. doi:10.1039/c1jm13640b.

[2] M. Tchernycheva, P. Lavenus, H. Zhang, A. V. Babichev, G. Jacopin, M. Shahmohammadi, F.H. Julien, R. Ciechonski, G. Vescovi, O. Kryliouk, InGaN/GaN Core–Shell Single Nanowire Light Emitting Diodes with Graphene-Based P-Contact, Nano Lett. 14 (2014) 2456–2465. doi:10.1021/nl5001295.

[3] R.R. Nair, P. Blake, A.N. Grigorenko, K.S. Novoselov, T.J. Booth, T. Stauber, N.M.R. Peres, A.K. Geim, Fine structure constant defines visual transparency of graphene., Science. 320 (2008) 1308. doi:10.1126/science.1156965.

[4] S. V. Morozov, K.S. Novoselov, M.I. Katsnelson, F. Schedin, D.C. Elias, J.A. Jaszczak, A.K. Geim, Giant intrinsic carrier mobilities in graphene and its bilayer, Phys. Rev. Lett. 100 (2008). doi:10.1103/PhysRevLett.100.016602.

[5]     C. Lee, X. Wei, J.W. Kysar, J. Hone, Measurement of the elastic properties and intrinsic strength of monolayer graphene., Science. 321 (2008) 385–388. doi:10.1126/science.1157996.

[6]     S. Tongay, M. Lemaitre, X. Miao, B. Gila, B.R. Appleton, A.F. Hebard, Rectification at Graphene-Semiconductor Interfaces: Zero-Gap Semiconductor-Based Diodes, Phys. Rev. X. 2 (2012) 011002. doi:10.1103/PhysRevX.2.011002.

[7]     A. Di Bartolomeo, Graphene Schottky diodes: An experimental review of the rectifying graphene/semiconductor heterojunction, Phys. Rep. 606 (2016) 1–58. doi:10.1016/j.physrep.2015.10.003.

[8]     G. Fan, H. Zhu, K. Wang, J. Wei, X. Li, Q. Shu, N. Guo, D. Wu, Graphene/silicon nanowire schottky junction for enhanced light harvesting, ACS Appl. Mater. Interfaces. 3 (2011) 721–725. doi:10.1021/am1010354.

[9]     H. Park, S. Chang, X. Zhou, J. Kong, T. Palacios, S. Gradečak, Flexible Graphene Electrode-Based Organic Photovoltaics with Record-High Efficiency, Nano Lett. 14 (2014) 5148–5154. doi:10.1021/nl501981f.

[10]    A. V. Babichev, H. Zhang, P. Lavenus, F.H. Julien, A.Y. Egorov, Y.T. Lin, L.W. Tu, M. Tchernycheva, GaN nanowire ultraviolet photodetector with a graphene transparent contact, Appl. Phys. Lett. 103 (2013) 201103. doi:10.1063/1.4829756.

[11]    G. Jo, M. Choe, C.-Y. Cho, J.H. Kim, W. Park, S. Lee, W.-K. Hong, T.-W. Kim, S.-J. Park, B.H. Hong, Y.H. Kahng, T. Lee, Large-scale patterned multi-layer graphene films as transparent conducting electrodes for GaN light-emitting diodes, Nanotechnology. 21 (2010) 175201. doi:10.1088/0957-4484/21/17/175201.

[12]    S. Chandramohan, J.H. Kang, B.D. Ryu, J.H. Yang, S. Kim, H. Kim, J.B. Park, T.Y.

Kim, B.J. Cho, E.-K. Suh, C.-H. Hong, Impact of Interlayer Processing Conditions on the Performance of GaN Light-Emitting Diode with Specific NiO x /Graphene Electrode, ACS Appl. Mater. Interfaces. 5 (2013) 958–964. doi:10.1021/am3026079.

[13]  A.K. Ranade, R.D. Mahyavanshi, P. Desai, M. Kato, M. Tanemura, G. Kalita, Ultraviolet light induced electrical hysteresis effect in graphene-GaN heterojunction, Appl. Phys. Lett. 114 (2019) 151102. doi:10.1063/1.5084190.

[14]  C.-J. Lee, S.-B. Kang, H.-G. Cha, C.-H. Won, S.-K. Hong, B.-J. Cho, H. Park, J.-H. Lee, S.-H. Hahm, GaN metal–semiconductor–metal UV sensor with multi-layer graphene as Schottky electrodes, Jpn. J. Appl. Phys. 54 (2015) 06FF08. doi:10.7567/JJAP.54.06FF08.

[15]  Y. Zhao, G. Wang, H.-C. Yang, T.-L. An, M.-J. Chen, F. Yu, L. Tao, J.-K. Yang, T.-B. Wei, R.-F. Duan, L.-F. Sun, Direct growth of graphene on gallium nitride by using chemical vapor deposition without extra catalyst, Chinese Phys. B. 23 (2014) 096802. doi:10.1088/1674-1056/23/9/096802.

[16]  T. Ciuk, I. Pasternak, A. Krajewska, J. Sobieski, P. Caban, J. Szmidt, W. Strupinski, Properties of Chemical Vapor Deposition Graphene Transferred by High-Speed Electrochemical Delamination, J. Phys. Chem. C. 117 (2013) 20833–20837. doi:10.1021/jp4032139.

[17]  J. Kierdaszuk, P. Kaźmierczak, A. Drabińska, K. Korona, A. Wołoś, M. Kamińska, A. Wysmołek, I. Pasternak, A. Krajewska, K. Pakuła, Z.R. Zytkiewicz, Enhanced Raman scattering and weak localization in graphene deposited on GaN nanowires, Phys. Rev. B. 92 (2015) 195403. doi:10.1103/PhysRevB.92.195403.

[18]  W. Liu, H. Li, C. Xu, Y. Khatami, K. Banerjee, Synthesis of high-quality monolayer and bilayer graphene on copper using chemical vapor deposition, Carbon N. Y. 49

(2011) 4122–4130. doi:10.1016/j.carbon.2011.05.047.

[19]   R. Fates, H. Bouridah, J.-P. Raskin, Probing carrier concentration in gated single, bi- and tri-layer CVD graphene using Raman spectroscopy, Carbon N. Y. 149 (2019) 390–399. doi:10.1016/j.carbon.2019.04.078.

[20]   J. Kierdaszuk, M. Tokarczyk, K.M. Czajkowski, R. Bożek, A. Krajewska, A. Przewłoka, W. Kaszub, M. Sobanska, Z.R. Zytkiewicz, G. Kowalski, T.J. Antosiewicz, M. Kamińska, A. Wysmołek, A. Drabińska, Surface-enhanced Raman scattering in graphene deposited on Al Ga1-N/GaN axial heterostructure nanowires, Appl. Surf. Sci. 475 (2019) 559–564. doi:10.1016/j.apsusc.2019.01.040.

[21]   H. Zhong, K. Xu, Z. Liu, G. Xu, L. Shi, Y. Fan, J. Wang, G. Ren, H. Yang, Charge transport mechanisms of graphene/semiconductor Schottky barriers: A theoretical and experimental study, J. Appl. Phys. 115 (2014) 013701. doi:10.1063/1.4859500.

[22]   S. Kim, T.H. Seo, M.J. Kim, K.M. Song, E.-K. Suh, H. Kim, Graphene-GaN Schottky diodes, Nano Res. 8 (2015) 1327–1338. doi:10.1007/s12274-014-0624-7.

[23]   A. Das, S. Pisana, B. Chakraborty, S. Piscanec, S.K. Saha, U. V Waghmare, K.S. Novoselov, H.R. Krishnamurthy, A.K. Geim, A.C. Ferrari, A.K. Sood, Monitoring dopants by Raman scattering in an electrochemically top-gated graphene transistor., Nat. Nanotechnol. 3 (2008) 210–215. doi:10.1038/nnano.2008.67.

[24]   T.M.G. Mohiuddin, A. Lombardo, R.R. Nair, A. Bonetti, G. Savini, R. Jalil, N. Bonini, D.M. Basko, C. Galiotis, N. Marzari, K.S. Novoselov, A.K. Geim, A.C. Ferrari, Uniaxial strain in graphene by Raman spectroscopy: G peak splitting, Grüneisen parameters, and sample orientation, Phys. Rev. B - Condens. Matter Mater. Phys. 79 (2009) 205433. doi:10.1103/PhysRevB.79.205433.

[25] J. Zabel, R.R. Nair, A. Ott, T. Georgiou, A.K. Geim, K.S. Novoselov, C. Casiraghi, Raman spectroscopy of graphene and bilayer under biaxial strain: Bubbles and balloons, Nano Lett. 12 (2012) 617–621. doi:10.1021/nl203359n.

[26] J. Yan, Y. Zhang, P. Kim, A. Pinczuk, Electric Field Effect Tuning of Electron-Phonon Coupling in Graphene, Phys. Rev. Lett. 98 (2007) 166802. doi:10.1103/PhysRevLett.98.166802.

[27] S. Pisana, M. Lazzeri, C. Casiraghi, K.S. Novoselov, A.K. Geim, A.C. Ferrari, F. Mauri, Breakdown of the adiabatic Born–Oppenheimer approximation in graphene, Nat. Mater. 6 (2007) 198–201. doi:10.1038/nmat1846.

[28] P.K. Ang, W. Chen, A.T.S. Wee, K.P. Loh, Solution-Gated Epitaxial Graphene as pH Sensor, J. Am. Chem. Soc. 130 (2008) 14392–14393. doi:10.1021/ja805090z.

[29] J. Binder, J.M. Urban, R. Stepniewski, W. Strupinski, A. Wysmolek, In situ Raman spectroscopy of the graphene/water interface of a solution-gated field-effect transistor: electron–phonon coupling and spectroelectrochemistry, Nanotechnology. 27 (2016) 045704. doi:10.1088/0957-4484/27/4/045704.

[30] V. Mišeikis, S. Marconi, M.A. Giambra, A. Montanaro, L. Martini, F. Fabbri, S. Pezzini, G. Piccinini, S. Forti, B. Terrés, I. Goykhman, L. Hamidouche, P. Legagneux, V. Sorianello, A.C. Ferrari, F.H.L. Koppens, M. Romagnoli, C. Coletti, Ultrafast, Zero-Bias, Graphene Photodetectors with Polymeric Gate Dielectric on Passive Photonic Waveguides, ACS Nano. 14 (2020) 11190–11204. doi:10.1021/acsnano.0c02738.

[31] K. Xu, H. Lu, E.W. Kinder, A. Seabaugh, S.K. Fullerton-Shirey, Monolayer Solid-State Electrolyte for Electric Double Layer Gating of Graphene Field-Effect Transistors, ACS Nano. 11 (2017) 5453–5464. doi:10.1021/acsnano.6b08505.

[32] D. Fuchs, B.C. Bayer, T. Gupta, G.L. Szabo, R.A. Wilhelm, D. Eder, J.C. Meyer, S. Steiner, B. Gollas, Electrochemical Behavior of Graphene in a Deep Eutectic Solvent, ACS Appl. Mater. Interfaces. 12 (2020) 40937–40948. doi:10.1021/acsami.0c11467.

[33] I. Pasternak, a. Krajewska, K. Grodecki, I. Jozwik-Biala, K. Sobczak, W. Strupinski, Graphene films transfer using marker-frame method, AIP Adv. 4 (2014) 097133. doi:10.1063/1.4896411.

[34] A. Di Bartolomeo, F. Giubileo, G. Luongo, L. Iemmo, N. Martucciello, G. Niu, M. Fraschke, O. Skibitzki, T. Schroeder, G. Lupina, Tunable Schottky barrier and high responsivity in graphene/Si-nanotip optoelectronic device, 2D Mater. 4 (2016) 015024. doi:10.1088/2053-1583/4/1/015024.

[35] J.M. Urban, P. Dąbrowski, J. Binder, M. Kopciuszyński, A. Wysmołek, Z. Klusek, M. Jałochowski, W. Strupiński, J.M. Baranowski, Nitrogen doping of chemical vapor deposition grown graphene on 4H-SiC (0001), J. Appl. Phys. 115 (2014) 233504. doi:10.1063/1.4884015.

[36] V. Bayot, L. Piraux, J.-P. Michenaud, J.-P. Issi, M. Lelaurain, A. Moore, Two-dimensional weak localization in partially graphitic carbons, Phys. Rev. B. 41 (1990) 11770–11779. doi:10.1103/PhysRevB.41.11770.

[37] J. Borysiuk, J. Sołtys, J. Piechota, Stacking sequence dependence of graphene layers on SiC (0001−)—Experimental and theoretical investigation, J. Appl. Phys. 109 (2011) 093523. doi:10.1063/1.3585829.

[38] L.M. Malard, M.A. Pimenta, G. Dresselhaus, M.S. Dresselhaus, Raman spectroscopy in graphene, Phys. Rep. 473 (2009) 51–87. doi:10.1016/j.physrep.2009.02.003.

[39] A.C. Ferrari, J.C. Meyer, V. Scardaci, C. Casiraghi, M. Lazzeri, F. Mauri, S. Piscanec,

D. Jiang, K.S. Novoselov, S. Roth, A.K. Geim, Raman spectrum of graphene and graphene layers, Phys. Rev. Lett. 97 (2006) 187401. doi:10.1103/PhysRevLett.97.187401.

[40] C. Cong, J. Jung, B. Cao, C. Qiu, X. Shen, A. Ferreira, S. Adam, T. Yu, Magnetic oscillation of optical phonon in ABA- and ABC-stacked trilayer graphene, Phys. Rev. B. 91 (2015) 235403. doi:10.1103/PhysRevB.91.235403.

[41] C. Cong, T. Yu, K. Sato, J. Shang, R. Saito, G.F. Dresselhaus, M.S. Dresselhaus, Raman Characterization of ABA- and ABC-Stacked Trilayer Graphene, ACS Nano. 5 (2011) 8760–8768. doi:10.1021/nn203472f.

[42] C.H. Lui, Z. Li, Z. Chen, P. V. Klimov, L.E. Brus, T.F. Heinz, Imaging Stacking Order in Few-Layer Graphene, Nano Lett. 11 (2011) 164–169. doi:10.1021/nl1032827.

[43] J. Yan, T. Villarson, E.A. Henriksen, P. Kim, A. Pinczuk, Optical phonon mixing in bilayer graphene with a broken inversion symmetry, Phys. Rev. B. 80 (2009) 241417. doi:10.1103/PhysRevB.80.241417.

[44] D.L. Mafra, P. Gava, L.M. Malard, R.S. Borges, G.G. Silva, J.A. Leon, F. Plentz, F. Mauri, M.A. Pimenta, Characterizing intrinsic charges in top gated bilayer graphene device by Raman spectroscopy, Carbon N. Y. 50 (2012) 3435–3439. doi:10.1016/j.carbon.2012.03.006.

[45] A. Drabińska, Photo- and Electroreflectance Spectroscopy of Low-Dimensional III-Nitride Structures, Acta Phys. Pol. A. 104 (2003) 149–164. doi:10.12693/APhysPolA.104.149.

[46] A. Drabinska, K.P. Korona, K. Pakula, J.M. Baranowski, Electroreflectance and photoreflectance spectra of tricolor III-nitride detector structures, Phys. Status Solidi.

204 (2007) 459–465. doi:10.1002/pssa.200673965.

[47] K.P. Korona, A. Drabińska, P. Caban, W. Strupiński, Tunable GaN/AlGaN ultraviolet detectors with built-in electric field, J. Appl. Phys. 105 (2009) 083712. doi:10.1063/1.3110106.

[48] K.P. Korona, A. Drabińska, A. Trajnerowicz, R. Bożek, K. Pakuła, J.M. Baranowski, Tuning of Spectral Sensitivity of AlGaN/GaN UV Detector, Acta Phys. Pol. A. 103 (2003) 675–681. doi:10.12693/APhysPolA.103.675.

[49] D.E. Aspnes, Band nonparabolicities, broadening, and internal field distributions: The spectroscopy of Franz-Keldysh oscillations, Phys. Rev. B. 10 (1974) 4228–4238. doi:10.1103/PhysRevB.10.4228.

[50] H. Shen, F.H. Pollak, Generalized Franz-Keldysh theory of electromodulation, Phys. Rev. B. 42 (1990) 7097–7102. doi:10.1103/PhysRevB.42.7097.

[51] A.P. Herman, L. Janicki, H.S. Stokowski, M. Rudzinski, E. Rozbiegala, M. Sobanska, Z.R. Zytkiewicz, R. Kudrawiec, Determination of Fermi Level Position at the Graphene/GaN Interface Using Electromodulation Spectroscopy, Adv. Mater. Interfaces. 7 (2020) 2001220. doi:10.1002/admi.202001220.

[52] L. Wang, M.I. Nathan, T. Lim, M.A. Khan, Q. Chen, High barrier height GaN Schottky diodes: Pt/GaN and Pd/GaN, Appl. Phys. Lett. 68 (1996) 1267–1269. doi:10.1063/1.115948.

[53] M. Drechsler, D.M. Hofmann, B.K. Meyer, T. Detchprohm, H. Amano, I. Akasaki, Determination of the Conduction Band Electron Effective Mass in Hexagonal GaN, Jpn. J. Appl. Phys. 34 (1995) L1178–L1179. doi:10.1143/JJAP.34.L1178.